
\magnification=\magstep1
\hsize = 15 true cm
\vsize = 22 true cm
\hoffset = 0.7 true cm
\voffset = 1 true cm
\parindent=1truecm

\message{NOTE: Postscript file containing figure appended to tex file - see
comments at head of tex file.}


\font\suff=cmmi6
\font\small=cmmi9

\font\bbb=msbm10 

\def\del{\partial}

\def\half{ {\textstyle{1 \over 2}} }

\def\vp{ \varphi }
\def\vpb{ {\bar \varphi} }
\def\vpt{\varphi_t}

\def\A{ {\hbox{\suff A} } }

\def\I{ {\hbox{\suff I} } }
\def\J{ {\hbox{\suff J} } }
\def\K{ {\hbox{\suff K} } }
\def\AA{{\hbox{\small A}} }
\def\II{{\hbox{\small I}} }

\def\V{ {\hbox{\small V} } }
\def\Vs{\V\,\,^*}
\def\qs{ q^{*  \A} }
\def\ps{ p^*_\A }

\def\L{\Lambda}

\def\R{ {\hbox{\bbb R} } }

\def\T{ {\scriptscriptstyle{\rm T} } }
\def\inv{ {\scriptscriptstyle{-1} } }

\def\zh{ {\hat z} }
\def\oh{ {\hat \omega} }

\def\G{ \Gamma}
\def\Gp{ \Gamma^* }
\def\Ge{ \bar \Gamma }
\def\Gpe{ \bar \Gamma^* }

\def\Xt{X_t}
\def\Xb{\bar X}

\def\z{ z^\mu }
\def\v*{ {v^*} }
\def\os{ \omega^* }
\def\mub{ {\bar \mu} }

\def\p{\prime}
\def\pp{ {\prime \prime} }

\def\wdg{ \wedge }

\def\d{ {\rm d} }


\def\npb{ \hbox{Nucl.} \hbox{Phys.} }
\def\plt{ \hbox{Phys.} \hbox{Lett.} }
\def\prd{ \hbox{Phys.} \hbox{Rev.} }

\def\prs{ \hbox{Proc.} \hbox{Roy.} \hbox{Soc.} }

\def\rmp{ \hbox{Rev.} \hbox{Mod.} \hbox{Phys.} }


\line{hep-th/9208009\hfill OUTP-92-11P}
\vskip 60pt

\centerline{A GEOMETRICAL APPROACH TO TIME-DEPENDENT GAUGE-FIXING}

\vskip 2truecm

\centerline{
Jonathan M. Evans\footnote*{Supported by SERC
Post-doctoral Research Fellowship B/90/RFH/8856}
}
\centerline{Theoretical Physics, University of Oxford}
\centerline{ 1 Keble Road, Oxford OX1 3NP, U.K.}
\vskip 5pt
\centerline{ and }
\vskip 5pt
\centerline{ Philip A. Tuckey}
\centerline{School of Physics and Materials, Lancaster University}
\centerline{Lancaster LA1 4YB, U.K.}
\vskip 2truecm

\centerline{ABSTRACT}

\vskip 10pt

\hsize=13.5truecm
\leftskip1.5truecm
\parindent=0truecm

When a Hamiltonian system is subject to constraints which depend
explicitly on time, difficulties can arise
in attempting to reduce the system to its physical phase space.
Specifically, it is non-trivial to restrict the system in such a way that
one can find a Hamiltonian time-evolution equation
involving the Dirac bracket.
Using a geometrical formulation, we derive an explicit condition
which is both necessary and sufficient for this to be possible, and we give a
formula defining the resulting Hamiltonian function.
Some previous results are recovered as special cases.

\hsize=15truecm
\leftskip0truecm
\parindent=1truecm

\vfill \eject

In systems such as string theory or general relativity
for which arbitrary time reparametrizations are symmetries,
any complete gauge fixing must involve the
imposition of conditions which are explicitly dependent on time.
(Here `time' means the evolution parameter entering in a canonical
formulation of the system.)
It was pointed out in [1] that the extension of the usual techniques
first developed by Dirac [2-5] to the case of general time-dependent
gauges is far from straightforward.
A careful analysis of the resulting problems was given and
some partial solutions were offered. In this sequel we show that a
geometrical approach yields a complete solution to the problem in a
sense made precise below.
We begin by summarizing the problem in conventional, non-geometrical
terms.

Consider a dynamical system consisting of (i) a phase space $\Gamma$
which can be parametrized locally by coordinates $\{ \z \}$ where $\mu
= 1 , \ldots , 2d$
and which is equipped with a Poisson bracket, to be denoted by square brackets;
(ii) a Hamiltonian function $H$ (possibly time-dependent) on $\Gamma$; and
(iii) a set of second-class constraints $\psi^i (\z , t)$ where $i =
1, \ldots , 2n$ which define
the {\it physical phase space\/} $\Gp \subset \G$ by means of the
local equations $\psi^i = 0$.
The dynamics is specified by Hamilton's equation
$$
{\d f \over \d t} = {\del f \over \del t} + [ f , H ]
\eqno(1) $$
for any function $f ( \z , t)$.
For consistency, the Hamiltonian $H$ must be such that the
constraints are preserved in time
$$
{\d \psi^i  \over \d t } =
{\del \psi^i \over \del t} + [ \psi^i , H ]  =  0 \quad {\rm when}
\quad \psi^j = 0 \ .
\eqno(2) $$
We shall consider throughout systems with
purely second-class constraints so that the matrix $c$ with elements
$$
c^{ij} = [ \psi^i , \psi^j ]
\eqno (3) $$
is non-singular. The situation dealt with in
[1] of a completely gauge-fixed system
which initially possesses only time-independent, first-class
constraints then arises as a special case.

By virtue of (2), any
trajectory which begins in $\Gp$ stays in $\Gp$ for all time, so we can
attempt to re-formulate the dynamics intrinsically on the
physical phase space.
The first step is to define the {\it Dirac
bracket\/} of any pair of functions $f$ and $g$ on $\G$ by
$$
[ f , g ]^* = [ f , g ] \, - \,
[ f , \psi^i ] \, (c^\inv)_{ij} \, [ \psi^j , g ]  \ .
\eqno (4) $$
This obeys $[ f , \psi^i ]^* = 0$ by construction and
therefore induces a well-defined bracket on $\Gp$ by restriction.
Although strictly the bracket on $\G$ and the induced bracket on $\Gp$ are
different entities, we shall also refer to the latter as the Dirac
bracket and we shall then understand the definition (4) to be supplemented by
the condition $\psi^i = 0$.
Note that the time dependence inherent in the definition of $\Gp
\subset \G$ does not influence the definition of the Dirac bracket in any way.

To proceed we must choose functions $\xi^a(\z,t)$ with $a = 1,
\ldots , 2d {-} 2n$ which will provide us with local coordinates on $\Gp$
and which we shall call {\it physical variables} following [1].
More precisely, these functions must define a smooth change of coordinates
$\{ \z \} \leftrightarrow \{\xi^a, \psi^i\}$ on $\G$ such that the
quantities $\{ \xi^a \}$ parametrize $\Gp$ on setting $\psi^i=0$.
To complete the gauge-fixing procedure we would like to write down a
Hamilton's equation on $\Gp$ using the new bracket (4).
In other words, we would like to find a Hamiltonian $H^*$ (possibly
time-dependent) on $\Gp$ such that
$$
{\d f \over \d t} = {\del f \over \del t} + [ f , H^* ]^*
\eqno(5)$$
for any function $f(\xi^a , t)$. Notice that here
$\del/\del t$ is defined with our chosen coordinates $\{\xi^a\}$ on
$\Gp$ held fixed, whereas in (1) the original coordinates $\{\z\}$
are held fixed.
Now the time dependence in the definition of $\Gp \subset \G$ is crucial:
since the change of coordinates $\{ \z \} \leftrightarrow \{\xi^a ,
\psi^i\}$ on $\G$ involves time explicitly
we must allow $H^*$ to differ from $H$
(or more accurately from the restriction of $H$ to $\Gp$).

It turns out that it is not always possible to find such a Hamiltonian for a
given choice of physical variables, although it was shown in [1] how
solutions could be obtained under certain assumptions on the gauge-fixing
conditions. In this paper we solve the general problem by
deriving necessary and sufficient
conditions on the choice of physical variables $\xi^a (\z , t)$
for the existence of a Hamiltonian $H^*$, as well as giving a formula defining
$H^*$.
The key is the use of geometrical methods which, as anticipated in
[1], allow a much clearer formulation of the problem.
The results of [1] will be recovered as special cases.

We must stress the importance of an equation of type (5).
At the purely classical level, a system
whose time evolution cannot be described in this way falls outside the realm
of conventional Hamiltonian mechanics.
Such an equation is also crucial in passing to the corresponding
quantum theory, where it becomes the Heisenberg equation of motion
and where its structure guarantees the existence of a
unitary time evolution operator.
If, in the absence of such an equation,
one tries to specify the quantum dynamics in some other fashion
then great care must be taken to ensure consistency.
Such an alternative scheme has been proposed by Gitman and Tyutin [5].

It is also shown in [5] that for any set of time-dependent
constraints there exists some canonical transformation to new
variables such that some subset of these is equivalent to the original
set of constraints.
In this sense one can in principle always remove any
time dependence from the constraints, but in practice the required
canonical transformation is usually very difficult to find.
The question which we address and solve here is quite different and
concerns the existence of a Hamilton's equation (5) for a
{\it prescribed\/} set of physical variables. This is
partly motivated by the fact that one often has strong physical
prejudices as to how the physical variables should be chosen.

The relativistic point particle provides perhaps the simplest example
of time-dependent gauge fixing; it is treated in detail in [1,5,8].
Examples of time-dependent gauge fixing in string theory
can be found in [6]. Discussions of this issue in
general relativity, and of the role of time in general, can be found in
[7] and references therein.

We now begin to translate our problem into the language of
symplectic geometry, building up the necessary vocabulary in stages.
For a dynamical system {\it without
constraints\/} [9] the phase space $\G$ is taken to be a {\it
symplectic manifold\/} with coordinates $\{\z\}$.
This means that $\G$ comes equipped with a {\it symplectic\/}, or
non-degenerate and closed, two-form $\omega$.
Let $\omega^{-1}$ be the corresponding antisymmetric contravariant
tensor so that the components of these tensors
are mutually inverse antisymmetric matrices.
The Poisson bracket of two functions on $\G$ is defined by
$$
[ f , g ] =  - \omega^{-1} ( \d f , \d g ) = - ( \omega^{-1} )^{\mu
\nu} \, {\del f \over \del z^\mu} \, { \del g \over \del z^\nu}
\quad\hbox{where}\quad
\omega = \half \, \omega_{\mu \nu} \, \d z^\mu \wdg \d z^\nu \ .
\eqno (6) $$
This bracket is clearly antisymmetric and also satisfies the Jacobi
identity because $\omega$ is closed.
Darboux's Theorem states that locally on $\G$ there exist coordinates
$\{ z^\mu \} = \{ q^m , p_m \}$ with $\omega = \d q^m \! \wdg \d p_m$
and the expression in (6) then takes on the form familiar from
non-geometrical treatments.

The time evolution of the system can be concisely specified as
follows.
Consider some trajectory on $\G$ which is parametrized by time $t$ and
which has tangent vector $v$ so that in components we can write
$$
\z = \gamma^\mu (t) \ ,
\qquad
v^\mu = \d \gamma^\mu / \d t \ ,
\eqno (7) $$
say. This trajectory is a solution of Hamilton's equation (1) precisely when
$$
i(v) \, \omega = \d H \ .
\eqno (8) $$
Here $i(v)$ denotes interior multiplication of a form by the vector
field $v$ so that in components
$(i(v)\, \omega)_\nu = v^\mu \omega_{\mu \nu}$.
To prove that (8) is equivalent to (1), note that along such
a trajectory
$\d f / \d t - \del f / \del t
= i(v) \, \d f = - \, \omega^{-1} ( \d f , i(v) \, \omega ) =  -
\omega^{-1} (\d f , \d H) = [ f , H ]$ where the first and second
equalities are identities, the third follows from (8) and the last
follows from the definition (6).

A related approach, which will prove more useful for our purposes,
involves the introduction of {\it extended phase space\/}.
This is a manifold $\Ge = \G {\times} \R$ with coordinates $\{z^\mub\} =
\{\z,z^0\!=\!t\}$, where $\{\z\}$ are coordinates on $\G$
and $t\in\R$ is time. The {\it Poincar\'e-Cartan two-form\/} on
$\Ge$ is defined by
$$
\Omega = \omega + \d H \wdg \d t \ ,
\eqno (9) $$
where $\omega$ is now to be understood as a form on $\Ge$ in the obvious way.
Using $\Omega$, Hamilton's equation can be written even more compactly.
Consider a trajectory on $\Ge$ which is a function of some auxiliary
parameter $s$ and which has tangent vector $\V$ so that
$$
z^\mub = {\bar \gamma}^\mub (s) \ , \qquad
{\V\,}^\mub = \d {\bar \gamma}^\mub / \d s \ ,
\eqno(10)$$
say. Any trajectory on $\G$ parametrized by $t$ is clearly equivalent to such a
trajectory on $\Ge$ and Hamilton's equation (1) holds if and only if
$$
i(\V\,) \, \Omega = 0 \ .
\eqno (11) $$
This follows from (8) by observing that when the tangent vector $v$ of
a trajectory on $\G$ parametrized by $t$ is regarded
as a vector field on $\Ge$ in the obvious way, then $\V = (\d t / \d s)
(v + \del / \del t )$.

We now progress to the geometrical description of a dynamical system
with {\it time-independent constraints\/}.
We assume that the local constraint equations $\psi^i=0$ define a
submanifold $X \subset \G$. It is useful to define the {\it
physical phase space\/} $\Gp$ to be some abstract copy of this
subset, rather than defining it to be the subset itself as was done in
our introductory remarks. These spaces can be identified by
an embedding map
$$
\vp : \Gp \rightarrow \G
\eqno(12)$$
which is a diffeomorphism onto its image $X$.
Let $\os$ be the pull back to $\Gp$ of the symplectic form
$\omega$ on $\G$ via this embedding
so that with a choice of coordinates $\{ \z \}$ on $\G$ and $\{ \xi^a \}$
on $\Gp$ we have\footnote\dag{
Given a map between manifolds, we shall frequently regard
coordinates or functions on its image as depending on
coordinates on its domain, as in (13).}
$$
\os = \half \, \omega^*_{ab} \, \d \xi^a \wdg \d \xi^b \quad\hbox{where}\quad
\omega^*_{ab} = {\del z^\mu \over \del \xi^a} \, {\del z^\nu \over \del \xi^b}
\, \omega_{\mu \nu} \ .
\eqno (13) $$
Since the exterior derivative commutes with the pull back, $\os$ is
closed. One can also show that $\os$ is
nondegenerate if and only if $c^{ij} = [\psi^i , \psi^j]$ is
nondegenerate. We assume this to be the case and then $\os$
makes $\Gp$ a symplectic manifold with an associated bracket defined
just as in (6).
This is the geometrical definition of the Dirac bracket on $\Gp$ and
one can show that it is equivalent to the formula (4). Details are given in the
appendix.

Consider a trajectory on $\G$ with tangent vector $v$ as in (7).
Suppose now that this trajectory corresponds to (i.e.~is the image under $\vp$
of) a trajectory in $\Gp$
which has tangent vector $v^*$, so that we can also write
$$
\xi^a = \gamma^{* \, a} (t) \ ,  \qquad v^{* \, a} = \d \gamma^{* \, a} /
\d t \ ,
\eqno (14) $$
say. Applying the chain rule and using the fact that there is, by definition,
no
time dependence in the relationship between $\{ z^\mu \}$ and $\{
\xi^a \}$ we have
$$
v^\mu = {\del z^\mu \over \del \xi^a} \, v^{* \, a} \ .
\eqno (15) $$
It follows from (13) and (15) that the equation of motion (8) is
equivalent to
$$
i(\v*) \, \os = \d H
\eqno (16) $$
on the physical phase space $\Gp$. This says precisely that (5) holds
with $H^* {=} H$.

Similar considerations apply to the extended phase space
formalism when the constraints are time-independent. We define
the {\it extended physical phase space\/} to be
$\Gpe = \Gp {\times} \R$ with coordinates
$\{\xi^{\bar a} \}=\{\xi^a, \xi^0\!=\!t\}$ where $\{\xi^a\}$ are
coordinates on $\Gp$ and $t\in\R$ is time.
Given $\vp$ in (12) there is a natural associated embedding
$$
\vpb : \Gpe \rightarrow \Ge \ ,
\qquad
\vpb (x,t) = (\vp(x),t)
\eqno (17) $$
which is a diffeomorphism onto its image $X {\times} \R$.
The Poincar\'e-Cartan two-form $\Omega$ on $\Ge$ pulls back under
$\vpb$ to a form $\Omega^*$ on $\Gpe$ given by
$$
\Omega^* = \omega^* + \d H \wdg \d t \ ,
\eqno (18) $$
where $\os$, given by (13), is now to be interpreted as a form on $\Gpe$.
It is easy to show that for a trajectory lying in $\Gpe$ with
tangent vector $\Vs$ the equation of motion (11) is equivalent to
$$
i(\Vs) \, \Omega^* = 0 \ .
\eqno (19)$$
The fact that $\os$ has no $\d t$ component again implies (by comparison of
(13), (18), (19) with (6), (9), (11)) that (5) holds with $H^*=H$.

Finally we can analyze the case of interest in which the system has
explicitly {\it time-dependent constraints\/}. To begin we must
discuss carefully the various spaces which arise
in the problem. At each fixed $t$ we have an {\it instantaneous
physical phase space\/} $\Xt \subset \G$ defined by the constraints.
We assume that the constraints are such that the
$X_t$ are all diffeomorphic to one another and that
the collection of all these instantaneous physical phase spaces is a
submanifold $\Xb = \{ (x,t) : x \in \Xt \} \subset \Ge$.
As before it is convenient to introduce a standard copy $\Gp$ of each $X_t$
which we call simply the {\it physical phase space\/}.
With the assumptions above, we can find a family of embeddings
$$
\vpt : \Gp \rightarrow \G
\eqno(20)$$
parametrized {\it smoothly\/} by time, such that each $\vpt$ is a
diffeomorphism from $\Gp$ onto its image $\Xt$.
We also define
$\Gpe = \Gp {\times} \R$ to be the {\it extended physical phase space\/}
and we notice that the smooth family of embeddings $\vpt$ is equivalent to a
single embedding
$$
\vpb : \Gpe \rightarrow \Ge \ , \qquad \vpb (x,t) = (\vpt(x) ,t)
\eqno(21)$$
which acts as the identity on the time factor and which is a diffeomorphism
onto its image $\Xb$.
There exist many inequivalent ways to choose the related embeddings $\vpt$
and $\vpb$, and the important point is that each of these
constitutes a distinct way of
identifying $\Xb$ as a product manifold $\Gp {\times} \R$.
The situation is summarized in the figure.

We can now give a more precise description of the physical variables
which will clarify their significance.
Consider a fixed set of coordinates $\{ \xi^a \}$ on $\Gp$.
For some choice of the embedding $\vpb$ (or the family $\vpt$) we can
use the property that this map  is a diffeomorphism onto its image
$\Xb$ to express the coordinates on the physical phase space
as functions on $\Xb$. These
functions, suitably extended to some patch in $\Ge$,
are precisely the physical variables $\xi^a (z^\mu , t)$.
A choice of physical variables therefore corresponds to a
specific choice of the embedding $\vpb$
and hence to a specific way of diffeomorphically
identifying $\Xb$ as a product manifold $\Gp{\times}\R$.
Only the values of the physical variables on $\Xb$
are important in this respect, since all suitably smooth extensions to
$\Ge$ clearly define the same identification.

In order to identify $\Xb$ as a product manifold globally, and not
just locally, the physical variables
$\xi^a (\z , t)$ chosen in each coordinate patch must be related by
time-independent transformations of the coordinates $\{ \xi^a \}$ where these
patches overlap.
A related point is that the Hamiltonian $H^*$ in (5) is clearly
insensitive to such transformations of the
$\{\xi^a\}$, but that the Hamiltonian might cease to exist or at least
would require modification in general after a
time-{\it dependent\/} transformation of the set $\{ \xi^a \}$.
We shall elaborate on these issues later.
The above discussion should also be contrasted with the
case of time-independent constraints
in which there is a {\it preferred\/} way to identify
$\Xb$ as a product manifold because $\Xt = X$ is actually constant in
time.
This corresponds to the existence of
preferred embeddings $\vp$ and $\vpb$ in (12) and (17)
for which the associated physical
variables can be chosen as time-independent functions $\xi^a (\z)$.

We noted earlier that the time dependence of the constraints has
essentially no effect on the way the Dirac bracket is constructed.
In geometrical terms, although $\vpt$ depends on time we can
use it at each fixed $t$ to pull back the symplectic
form $\omega$ on $\G$ to a symplectic form $\os$ on
$\Gp$, which defines a bracket by an equation exactly like (6).
The formulas (13) for the components of $\os$
and (4) for the Dirac bracket are clearly unaltered. What is new,
however, is that both $\os$ and the Dirac bracket can now change with
time so that the physical phase space has a time-varying
symplectic structure. We shall return to this issue below.

We are now faced with the central question of how to describe the
dynamics. The important point is that the equation of motion (8)
no longer leads to an equation of the same form (16)
on the physical phase space as it did
in the time-independent case. One way to understand this is to
consider a trajectory in $\Gp$ and to use the coordinate expressions
given in (7) and (14). The chain rule tells us that the components of
the tangent vectors are now related by
$$
v^\mu = {\del z^\mu \over \del \xi^a} \, v^{* \, a} + {\del z^\mu
\over \del t} \ ,
\eqno (22) $$
instead of by (15), so that $v$ is no longer tangent to $\Xt$ in general.
The new inhomogeneous term complicates the structure of the resulting
equation on $\Gp$ and from this point of view
it is unclear how best to proceed.

Using the formalism of extended
phase space, however, the situation is simpler in one important respect.
Because time is reduced to the status of a coordinate on $\Ge$, with
trajectories being parametrized by an auxiliary quantity $s$, the time
dependence of the embedding $\vpb$ does not influence
the reduction of the dynamical equations to $\Gpe$.
We can therefore immediately write
$$
i(\Vs) \, \Omega^* = 0
\eqno (23) $$
as the correct equation of motion on $\Gpe$, where as before
$\Omega^*$ is the pull-back of $\Omega$ under $\vpb$.
But our problem now manifests itself in the fact that $\Omega^*$
no longer has the structure exhibited in (18) in general.
We find instead
$$
\Omega^* = \os \, + \, \d H \wdg \d t \, + \, A \wdg \d t
\eqno (24) $$
where $\os$ is given by (13) and where
$$
A = - \, {\del z^\mu \over \del t} \,
{\del z^\nu \over \del \xi^a} \,
\omega_{\mu \nu} \, \d \xi^a \ .
\eqno (25) $$
Note that $\os$ is the pull-back to $\Gp$ of $\omega$ on $\G$ using $\vpt$
whereas $\os+A\wdg\d t$ is the pull back to $\Gpe$ of $\omega$ on $\Ge$
using $\vpb$.
The decomposition of this last pulled-back form into two terms $\os$ and
$A\wdg\d t$ is clearly independent of the choice of the coordinates
$\{ \xi^a \}$ on $\Gp$, depending only on the chosen embeddings $\vpt$
and $\vpb$. The decomposition would change under a general
time-{\it dependent\/} coordinate transformation on $\Gpe$ however.

Once the problem is presented in this way, the solution is
straightforward. To have a Hamilton's equation (5)
it is necessary and sufficient that
$$
\Omega^* = \os \, + \, \d H^* \! \wdg \d t
\eqno (26) $$
by comparison with (18).
But from (24) this is true precisely when
$$
A = \d K \quad {\rm mod} \, \, \d t
\qquad \Rightarrow \qquad H^* = H + K
\eqno (27) $$
for some function $K$.
Here mod $\d t$ means that equality
holds up to terms proportional to $\d t$.
It is not difficult to prove that
the Poincar\'e Lemma [9] holds mod $\d t$, so that a form which
contains no $\d t$
terms (such as $A$) is closed mod $\d
t$ if and only if it is locally exact mod $\d t$. One therefore has
the following concrete condition
$$
\d A = 0  \quad {\rm mod} \, \, \d t \qquad \iff \qquad
{\del \over \del \xi^{ [b } }
\left \{
{\del z^\nu \over \del \xi^{ a ]} } \,
{\del z^\mu \over \del t} \,
\omega_{\mu \nu} \right \} = 0
\eqno (28) $$
for the existence locally of a Hamiltonian $H^*$, and an explicit expression
for
the Hamiltonian is then derivable from (27).
There is also a potential global obstruction if $A$ defines a
non-trivial cohomology class in H$\vphantom{H}^1(\Gp)$.
There are a number of simple situations where this can be ruled out --
if $\Gp$ is simply connected for example -- but a more detailed
discussion is beyond the scope of the present paper.

Equation (27) has one more important consequence.
We emphasized above that the time dependence of the constraints implies
that the symplectic form $\os$ and hence the
Dirac bracket on $\Gp$ will also be explicitly time-dependent in general.
There seem to be no obvious grounds for objecting to this, and it
might appear that we are forced to accept this new feature despite
the fact that it runs counter to our experience with
conventional classical mechanics.
It turns out, however, that if (27) holds then
$\os$ is automatically time-independent.
This can be deduced straightforwardly
by working in components, comparing the time derivative of $\os$ with
the condition (28) above and using the fact that $\os$ is closed on
$\Gp$.
Thus (28) is actually a necessary and sufficient condition both for
the existence of a Hamilton's equation (5) and also for the time
independence of the Dirac bracket occurring in this equation.

The problem we posed is now solved, but it is instructive nevertheless to
take the analysis one step further and to frame things more
systematically.
As we have formulated it here, the problem of specifying the dynamics
of a system in conjunction with time-dependent gauge
fixing is the problem of how the structure of the
Poincar\'e-Cartan two-form is affected by pulling back under the embedding
$$
\vpb : \, \Gp \times \R \, \rightarrow \, \G \times \R \ .
\eqno (29) $$
This embedding specifies a particular way of identifying the
submanifold $\Xb \subset \G \times \R $ with the product $ \Gp \times \R$
and it is defined in local coordinates by a particular choice of
physical variables.

Whenever a manifold has a product structure $M = M^\p {\times} M^\pp$ there
is an induced decomposition of the space of differential forms
of a given degree
$\L^p (M) = \oplus_{r {+} s {=} p} \, \L^{(r,s)}$ where
$\L^{(r,s)}= \L^{r} (M^\p) \otimes
\L^{s} (M^\pp)$, and an associated decomposition of the exterior
derivative $\d = \d^\p + \d^\pp$ where $\d^\p : \L^{(r,s)}
\rightarrow \L^{(r{+}1,s)}$ and $\d^\pp : \L^{(r,s)}
\rightarrow \L^{(r,s{+}1)}$.
For each of the extended phase spaces $\Ge$ and $\Gpe$ the time factor is
one-dimensional and so the decomposition of a general $p$-form
$\alpha$ involves just two terms; let us write
$\alpha = \alpha^\p + \alpha^\pp$ where $\alpha^\p \in
\L^{(p,0)}$ and $\alpha^\pp \in \L^{(p{-}1,1)}$.
Thus for the Poincar\'e-Cartan form on $\Ge$ we have
$$
\Omega^\p = \omega \ , \qquad
\Omega^\pp = \d H \wdg \d t \ ,
\eqno (30) $$
while for its pull-back to $\Gpe$ we have
$$
\Omega^{* \, \p} = \os \ , \qquad
\Omega^{* \, \pp} = (A + \d H) \wdg \d t \ .
\eqno (31) $$

The key property of $\Omega$ which ensures
that (11) leads to Hamilton's equation (1)
is that it has the structure exhibited in (30) above
with $\d^\p \omega = 0$.
This property can be exactly
characterized locally by saying that both
$\Omega^\p$ and $\Omega^\pp$ are separately closed with respect to
$\d^\p$ on $\Ge$ (using the generalized Poincar\'e Lemma mentioned above).
Now the criterion for the existence of a Hamilton's equation (5) on
physical phase space
can be similarly expressed as the condition that both
$\Omega^{* \, \p}$ and $\Omega^{* \, \pp}$ are separately
closed with respect to $\d^\p$ on $\Gpe$:
$$\eqalignno{
\d^\p \Omega^{* \, \p} & = 0
& (32a) \cr
\d^\p \Omega^{* \, \pp} & = 0 \ .
& (32b) \cr
} $$
But since $\omega$ in (9) is independent
of time, we know also that $ \d \Omega = 0$ implying $\d \Omega^* = 0$
which is equivalent to
$$\eqalignno{
\d^\p \Omega^{* \, \p} & = 0
& (33a) \cr
\d^\pp \Omega^{* \, \p} + \d^\p \Omega^{* \, \pp} & = 0 \ .
&(33b) \cr
}$$
Because these last equations hold automatically, only $(32b)$ actually
has any content and moreover it is equivalent to the condition
$$
\d^\pp \Omega^{* \, \p} = 0 \ .
\eqno (34) $$

To relate this to our previous work, it is convenient to
employ an abuse of notation and to write the exterior
derivatives on $\Gpe$ as
$\d^\p = \d \xi^a (\del / \del \xi^a)$ and
$ \d^\pp = \d t \, ( \del / \del t)$.
This allows us to read off, using (31), the content of equations (32-34).
Clearly $(32a)$ says just that $\os$ is closed on $\Gp$
and $(32b)$ yields the condition (28) for $A$ found earlier.
But we have also found that $(32b)$ is equivalent to (34)
which says precisely that $\os$ is independent of time, as discussed above.

We stated earlier that the physical variables $\xi^a (\z , t)$ defined
in each coordinate patch should be
related on overlaps by time-independent transformations of
the set $\{ \xi^a \}$ if we wish to identify $\Xb$ as a product
manifold globally.
But the subsequent analysis and the criterion (27) we derived for the existence
of a Hamilton's equation were purely local. We are therefore free to adopt a
more flexible approach and to look for
solutions to (27) independently in each coordinate patch.
Under these circumstances, the identification of $\Xb$ as a product
will hold only locally, being defined by
the particular choice of $\xi^a ( \z , t)$ specific to each patch.
This choice will also define, via $\d \xi^a$ and $\d t$, local decompositions
of the spaces of differential forms on each patch, as described above.
The $\xi^a (\z , t)$ will be related by time-dependent
transformations of the set $\{ \xi^a \}$ on overlaps and the
associated symplectic forms $\os$ and Hamiltonians $H^*$ will differ in these
regions. See also the remarks following (25).

It remains for us to demonstrate how the results of [1] can be
recovered within the present geometrical framework. This will also
serve to illustrate how the techniques we have developed work in practice.

We first treat case (B) of [1]. Suppose that on $\G$ we have coordinates
$\z = \{ q^m , p_m \} $ in which $\omega = \d q^m \! \wdg \d p_m $.
Let $\{ Q^\I , P_\I \}$ with $\II = 1 , \ldots , n$
and $\{ q^\A , p_\A\}$ with $\AA = 1, \ldots , d{-}n$
be disjoint subsets of these coordinates and suppose that the
constraints have the form
$\{ \psi^i \} = \{ \chi^\I , \phi^\I \}$ where
$$
\chi^\I = Q^\I - \zeta^\I (q^\A , t)
\eqno (35) $$
for certain functions $\zeta^\I$. Result (B) of [1] states that for
physical variables $\{ \xi^a \} = \{ \qs , \ps \}$ there exists a
Hamiltonian $H^*$ where
$$
\qs = q^\A \ , \qquad \ps = p_\A \, + \, { \del \zeta^\I \over \del q^\A
} \, P_\I \ , \qquad
H^* = H \, - \, { \del \zeta^\I \over \del t} \, P_\I
\eqno(36)$$
and this last expression is to be thought of as a function of the
physical variables.

To prove this we shall calculate the one-form $A$. This
illustrates one general practical approach to dealing with the physical
variables: we can take the explicit expressions $\xi^a ( \z , t )$
and the equations $\psi^i (\z ,t) = 0$ and solve them to express
$\z (\xi^a , t)$, thus reducing to the physical phase space.
For the case at hand we have
$$\eqalign{
&Q^\I = \zeta^\I( \qs , t ) \ , \qquad
P_\I  = \eta_\I ( \qs , \ps , t)
\cr
&q^\A = \qs \ , \qquad \qquad p_\A = \ps - {\del \zeta^\I \over \del
\qs} \, \eta_\I
\cr
}\eqno (37) $$
where the functions $\eta_\I$ depend in detail upon the entire set of
constraints.
Now because $( \del q^\A / \del t )_{\xi^a} = 0$ we have
$$ A =
{ \del p_\A \over \del t} \,  { \del q^\A \over \del \xi^a } \d \, \xi^a
\, - \, { \del \zeta^\I \over \del t } \, { \del \eta_\I \over \del \xi^a }
\, \d \xi^a
\, + \, { \del \eta_\I \over \del t} \, { \del \zeta^\I \over \del \xi^a }
\, \d \xi^a \ .
\eqno (38) $$
By making further use of (37), and in particular the fact that $ ( \del
\zeta^I / \del \ps)_{ \qs , t} \! = 0$, the first term in this expression for
$A$ can be written
$$
{\del p_\A \over \del t} \, { \del q^\A \over \del \xi^a } \, \d \xi^a
=
- { \del \over \del t}
\left \{
{\del \zeta^\I \over \del \qs } \, \eta_\I
\right \}
\d \qs
=
- {\del \over \del t}
\left \{
{\del \zeta^\I \over \del \xi^a } \, \eta_\I
\right \}
\d \xi^a \ .
\eqno (39) $$
Combining these expressions gives
$$
A =
- \, {\del \over \del \xi^a}
\left \{
{\del \zeta^\I \over \del t } \, \eta_\I
\right \}
\d \xi^a
=
- \, \d \left \{
{\del \zeta^\I \over \del t } \, \eta_\I
\right \}
\quad {\rm mod } \, \, \d t
\eqno (40) $$
and so, as claimed, the chosen physical variables admit a Hamiltonian
$$
H^* = H \, - \, { \del \zeta^\I \over \del t} \, \eta_\I \ .
\eqno (41) $$

The other case to be discussed is result (A) of [1].
We assume the constraints take the form
$\{ \psi^i \} = \{\chi^\I , \phi^\I \}$ where
$$ ( { \del \phi^\I / \del t}  )_{\z} = 0 \ ,
\qquad \quad
[ \phi^\I  , \phi^\J ] = 0 \quad {\rm when} \quad \phi^\K=0 \,.
\eqno (42) $$
The physical variables
$\xi^a (\z , t)$ are defined by the requirements that they are
time-independent and gauge-invariant
$$
( {\del \xi^a / \del t} )_{\z} = 0 \ ,
\qquad \quad
[ \phi^\I , \xi^a ] = 0 \quad {\rm when} \quad \phi^\J=0 \,.
\eqno (43)$$
Result (A) of [1] then states that (5) holds with $H^* = H$.

To prove this we use another approach of general applicability,
namely we consider a coordinate transformation $\{\z , t\} \rightarrow
\{\xi^a , \psi^i , t\}$ on $\Ge$, view (13), (25) and (28) as defined
on $\Ge$, then reduce to the physical phase space by setting
$\psi^i = 0$ (and then (28) has to hold only after so doing).
In the present case it is convenient to introduce as a shorthand
$ \{ \zh^\mu \} = \{ \xi^a , \chi^\I , \phi^\I \} \ $
and to let $\oh_{\mu \nu}$ denote the components of the symplectic
form in this coordinate system. Equation (6) tells us that
$[ \zh^\mu , \zh^\nu ] = - (\oh^{-1} )^{\mu \nu}$ and therefore
conditions (42) and (43) above imply that when
$\psi^i=0$ we have the block forms
$$
\oh^{-1} = \pmatrix{
a & x & 0 \cr
-x^\T & b & y \cr
0 & -y^\T & 0 \cr
}
\qquad \Rightarrow \qquad
\oh = \pmatrix{
\alpha & 0 & - \lambda^\T \cr
0 & 0 & - \mu^\T \cr
\lambda & \mu & \beta \cr
} \ .
\eqno (44) $$
Here $a$ and $b$ are antisymmetric, $a$ and $y$ are invertible, with
$$
\alpha = a^\inv , \qquad \mu = y^\inv , \qquad \lambda = y^\inv x^\T
a^\inv , \qquad \beta = y^\inv ( b + x^\T a^\inv x) (y^\T)^\inv \ ,
\eqno (45) $$
although actually only the block structure of these matrices is
important for our purposes.
Now when we regard the coordinates $\zh^\mu$ as functions of the
original coordinates $\{ \z,t \}$ on $\Ge$, only the quantities $\chi^\I$
depend on time explicitly. Consequently we have the identity
$$
\left ( \del z^\mu \over \del t \right )_{ \! \zh^\mu } + \, \,
\left ( {\del \chi^\I \over \del t} \right )_{\! z^\mu} \!
\left ( {\del z^\mu \over \del \chi^\I} \right )_{\! \xi^a , \phi^I , t}
\!   = 0
\eqno (46) $$
and on substituting this in (25) we find
$$
A = {\del \chi^\I \over \del t}
\left \{
{\del z^\mu \over \del \chi^\I} \, {\del z^\nu \over \del \xi^a }
\, \omega_{\mu \nu}
\right \} \d \xi^a \ .
\eqno (47)$$
But the factor in curly brackets is one of the block entries in
$\oh$ which vanishes when $\psi^i=0$ according to (44).
Hence $A=0$ on $\Gpe$ and the result is established.

In conclusion: we have derived an explicit condition (28) on a
set of physical variables $\xi^a (\z , t)$ which is necessary
and sufficient both for the existence (locally) of a Hamiltonian $H^*$ in (5)
--
which is then determined by (27) -- and for the constancy in time of the Dirac
bracket (4) on the physical phase space.
It follows from the work of
Gitman and Tyutin [5] that for {\it any\/} set of constraints there exists
{\it some\/} set of physical variables which satisfy these conditions.
The proof is of no practical help in finding these, however,
hence the utility of our result.
Note also that in the special case where there are {\it no\/} constraints, our
analysis determines when there exists a Hamilton's equation
for some new set of coordinates $\{\xi^\mu,t\}$ on $\Ge$ which are
related by a time-dependent transformation to the original set $\{\z,t\}$.

The geometrical techniques used here have proved far more efficient than the
original approach of [1]. There the emphasis was also placed on
finding an expression on the full phase space $\G$ which gave the
desired equation on restriction, rather than on working directly on the
physical phase space $\Gp$.
Such an approach can also be expressed in geometrical terms
although we have not pursued the details here.
In the future it would be interesting to apply our results to
specific examples such as those in [6,7].
It would also be very interesting to study the extension of this work
to the quantum case and particularly its relationship to geometric
quantization.
\vskip 5pt

PAT acknowledges extremely helpful conversations with David Hartley.
JME is grateful to the SERC for financial support.
PAT is grateful for the support of a BP Venture Research Fellowship.

\vskip 15pt

\centerline{APPENDIX: THE DIRAC BRACKET}
\vskip 10pt
\noindent
Here we show how the geometrical definition (6) is related to formula (4).
We can ignore all questions of time dependence and work throughout at
some fixed instant. Take coordinates
$\{ \zh^\mu \} = \{ \xi^a , \psi^i \}$ on $\G$ and let the components
of $\omega$ and
$\omega^\inv$ have the corresponding block forms
$$
- \oh = \pmatrix{
\alpha & - \beta^\T \cr
\beta & \gamma \cr
} , \qquad
- \oh^\inv = \pmatrix{
a & b \cr
- b^\T & c \cr
}
\eqno (A.1) $$
so that
$$
a \alpha + b \beta = 1 \ , \qquad - b^\T \alpha + c \beta = 0 \ ,
\qquad c \gamma + b^\T \beta^\T = 1 \ .
\eqno (A.2) $$
If $c$ is invertible the first two equations imply that $\alpha$ is
invertible with
$$
\alpha^\inv = a + b c^\inv b^\T \ ,
\eqno (A.3) $$
whilst if $\alpha$ is invertible the last two equations imply that $c$
is invertible with
$$
c^\inv = \gamma + \beta \alpha^\inv \beta^\T \ .
\eqno (A.4) $$
Now from (6) we have $- (\oh^{-1})^{\mu \nu} = [ \zh^\mu , \zh^\nu ]$
or in detail
$$
a^{ab} = [ \xi^a , \xi^b ] \ , \qquad
b^{a i} = [ \xi^a , \psi^i ] \ , \qquad
c^{ij} = [ \psi^i , \psi^j ] \ .
\eqno (A.5) $$
Also, taking $\{ \xi^a \}$ as coordinates on $\Gp$, the components
of $\os$ are
$$
\os_{ab} = - \alpha_{ab} \ .
\eqno (A.6) $$
The remarks above therefore prove that
$\os$ is
non-degenerate if and only if $[ \psi^i , \psi^j ]$ is non-degenerate,
as stated in the text. Furthermore, the
bracket on $\G$ defined by
$$
[f , g]^* = {\del f \over \del \xi^a} \, { \del g \over \del \xi^b}
\, (\alpha^\inv)^{ab}
\eqno (A.7) $$
clearly restricts to the bracket on $\Gp$ defined by (6) by virtue of
$(A.6)$. But $(A.3)$ and $(A.5)$ allow us to express the
right-hand side in terms of Poisson brackets:
$$\eqalign{
[f , g ]^* & = { \del f \over \del \xi^a} \, { \del g \over \del \xi^b} \,
( \, [ \xi^a , \xi^b ] \,
+ \, [\xi^a , \psi^i ] \, (c^{-1})_{ij} \, [\xi^b , \psi^j] \, )
\cr
& = [f , g ] \, - \, [f , \psi^i ] \, (c^{\inv})_{ij} \, [ \psi^j , g]
\cr
}\eqno(A.8) $$
which reproduces (4). The equivalence of the definitions is therefore
established.

\vfill \eject

\centerline{REFERENCES}
\vskip 10pt

\item{[1]} J.M. Evans, \plt {\bf B256} (1991) 245
\vskip 5pt

\item{[2]} P.A.M. Dirac, Canad. J. Math. {\bf 2} (1950) 129;
\prs {\bf A246} (1958) 326;
{\it Lectures on Quantum Mechanics}, Belfer Graduate School of Science, Yeshiva
University (Academic, New York, 1964)
\vskip 5pt

\item{[3]} A. Hanson, T. Regge and C. Teitelboim, {\it Constrained Hamiltonian
Systems}
(Accademia Nazionale dei Lincei, Rome, 1976)
\vskip 5pt

\item{[4]} K. Sundermeyer, {\it Constrained Dynamics}, Springer
Lecture Notes in Physics Vol. 169 (Springer-Verlag, Berlin-Heidelberg-New
York, 1982)
\vskip 5pt

\item{[5]} D.M. Gitman and I.V. Tyutin, {\it Quantization of Fields
with Constraints}
\hfil \break
(Springer-Verlag, Berlin-Heidelberg-New York, 1990)
\vskip 5pt

\item{[6]} P. Goddard, J. Goldstone, C. Rebbi and C.B. Thorn, \npb {\bf
B56} (1973) 109;
\hfil \break
J. Scherk, \rmp {\bf 47} (1975) 123;
\hfil \break
P. Goddard, A. Hanson and G. Ponzano, \npb {\bf B89} (1975) 76
\vskip 5pt

\item{[7]}
P.A.M. Dirac, \prd {\bf 114} (1959) 924;
\hfil \break
N. Manojlovi\'c and A. Mikovi\'c, {\it Gauge Fixing and
Independent Canonical Variables in the Ashtekar Formulation of General
Relativity}, Imperial College/Queen Mary and Westfield College preprint
Imperial/TP/90-91/18 \hfil\break QMW/PH/91/7;
\hfil \break
C.J. Isham, {\it Conceptual and Geometric Problems in
Quantum Gravity\/},
\hfil \break
Lectures at 1991 Schladming Winter School,
Imperial College preprint \hfil\break Imperial/TP/90-91/14
\vskip 5pt

\item{[8]} A. Ogielski and P.K. Townsend, Lett.~Nuovo Cim.~{\bf 25} no.2
(1979) 43
\vskip 5pt

\item{[9]} V.I. Arnol'd, {\it Mathematical Methods of Classical Mechanics}
(Springer-Verlag, New York, 1978)
\vskip 5pt

\item{[10]} G. Marmo, N. Mukunda and J. Samuel, Rev.~Nuovo Cim.~{\bf 6}
no.2 (1983) 1
\vfill\eject

\centerline{FIGURE CAPTION}
\vskip 20pt

Extended phase space is $\Ge = \G\times\R$. Extended physical phase space is
$\Gpe = \Gp\times \R$. $\vpb$ is an embedding from $\Gpe$ to $\Ge$ with
image $\Xb$. It is equivalent to the family of embedding maps
$\vpt$ from $\Gp$ to $\G$ with images $\Xt$.

\bye